%% file: paper.tex
\documentclass[lettersize,journal]{IEEEtran}
\usepackage{cite}
\usepackage{algorithmic}
\usepackage{algorithm}
\usepackage{array}
\usepackage{graphicx}
\usepackage[caption=false,font=normalsize,labelfont=sf,textfont=sf]{subfig}
\usepackage{url}
\usepackage{verbatim}
\usepackage{bm}
\usepackage{xcolor}
\usepackage{comment}
\usepackage{hyperref}
\usepackage[affil-it]{authblk}

\newcommand{\pg}[1]{}
\newcommand{\todo}[1]{}

\def\BibTeX{{\rm B\kern-.05em{\sc i\kern-.025em b}\kern-.08em
    T\kern-.1667em\lower.7ex\hbox{E}\kern-.125emX}}

\begin{document}

\title{Correct Wrong Path}
\author{Bhargav Reddy Godala\textsuperscript{\symbol{42}},  
Sankara Prasad Ramesh\textsuperscript{\symbol{42}}, 
Krishnam Tibrewala,  
Chrysanthos Pepi, 
Gino Chacon, 
Svilen Kanev,  
Gilles A. Pokam, 
Daniel A. Jiménez,
Paul V. Gratz, 
David I. August
\thanks{Bhargav Reddy Godala, Gino Chacon and Gilles A. Pokam  from Intel (\{bhargav.reddy.godala, ginoachacon, gilles.a.pokam\}@intel.com). Sankara Prasad Ramesh from Nvidia (spramesh@ucsd.edu). Krishnam Tibrewala, Chrysanthos Pepi, Daniel A. Jiménez, and Paul V. Gratz from Texas A\&M University (\{krishnamtibrewala, cpepis, djimenez, pgratz\}@tamu.edu). Svilen Kanev from Google (skanev@google.com). David I. August  from Princeton University (august@princeton.edu).

\symbol{42} Bhargav and Sankara Contributed Equally to this work.
}
}

\maketitle
\thispagestyle{plain}
\pagestyle{plain}


\input{abstract}
\input{introduction}


\bibliographystyle{IEEEtranS}
\bibliography{refs}

\end{document}

%% file: abstract.tex
\begin{abstract}

Modern OOO CPUs have very deep pipelines with large branch misprediction recovery 
penalties. Speculatively executed instructions on the
wrong path can significantly change cache state, depending on speculation levels.  Architects often 
employ trace-driven simulation models in the design exploration stage, which sacrifice precision for speed.
Trace-driven simulators are orders of magnitude faster than execution-driven models, reducing the often hundreds of thousands of simulation hours needed to explore new micro-architectural ideas.
Despite this strong benefit of trace-driven simulation, these often fail to adequately model the consequences of wrong path because obtaining them is nontrivial.
Prior works consider either a positive or negative impact of wrong path but not both.
Here, we examine wrong path execution in simulation results and design a set of infrastructure for enabling wrong-path execution in a trace driven simulator. Our analysis shows the wrong path affects structures on both the instruction and data sides extensively, resulting in performance variations ranging from $-3.05$\% to $20.9$\% when ignoring wrong path. To benefit the research community and enhance the accuracy 
of simulators, we opened our traces and tracing utility in the hopes that industry can provide wrong-path traces generated by their internal simulators, enabling academic simulation without exposing industry IP.

\end{abstract}

%% file: introduction.tex
\vspace{-20pt}
\section{Introduction}
Accurate modeling of modern Out-of-Order (OOO) CPUs is very expensive in terms of
simulator complexity and simulation times. Simulators facilitate prototyping of 
micro-architectures, but their complexity and simulation times can be significant
bottlenecks in design space exploration. Consequently, architects often use simpler, and
faster, though less accurate, trace-driven simulators, trading accuracy for speed.
Trace driven simulators do not model wrong path (WP), as traces are typically generated from the execution of a workload on a real machine and thus can only contain correct path, executed instructions.  Ignoring WP, however, can lead to incorrect estimation of performance.

Prior works~\cite{wrong_path_intel,wp_eip} have proposed solutions to model
either positive or negative impact but not both. Here, a positive impact is observed
when a cache line fetched in WP is later used by correct path instructions, thus acting like a prefetch.  A negative impact is observed when the cache is polluted with WP data or instructions that are never used.  The alternative
is to use more accurate execution-driven models, but they are orders of magnitude slower
than trace-driven simulators~\cite{gem5,ChampSim}. 

\begin{figure}[t]
    \centering
    \vspace{-10pt}
     \includegraphics[width=\columnwidth]{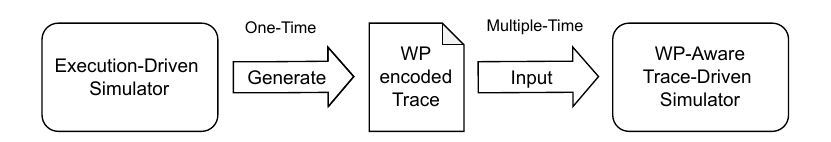}
    \vspace{-15pt}
    \caption{WP traces from execution-driven to trace-driven simulator}
    \vspace{-15pt}
    \label{fig:work_flow}
\end{figure}

Instead of using actual execution to generate traces, we propose to use an execution-driven simulator
to obtain traces.  This approach allows us to observe and annotate both the correct- and wrong-path instructions into the trace.
To facilitate this, we also propose a new trace format and a set of design changes to a
trace-driven simulator to conditionally model WP instructions and repair
state of the CPU before executing in the correct path. Fig~\ref{fig:work_flow} shows our proposed simulation flow where a one-time cost of running the execution-driven model enables quick WP-aware trace-based design space exploration.  Here we implemented our approach in
the widely used ChampSim~\cite{ChampSim} simulator, though this approach could be implemented in any trace-driven simulator.

This paper makes the following contributions:
\begin{itemize}
  \item Extensive analysis and metrics to measure the impact of WP instructions.
  \item A new tracing utility leveraging gem5 and a new WP trace format~\cite{gem5_champsim_tracer}.
  \item Correct WP modeling in a trace-driven simulator~\cite{champsim_wp}.
  \item An open set of traces of widely used data center and SPEC workloads~\cite{champsim-traces}.
\end{itemize}

\section{WP Model}

Simulating instructions in the WP involves first, knowing what instructions are down the WP after a branch misprediction. Then letting those instructions
flow from fetch till execute states of processor pipeline, repairing
all pipeline stages by flushing instructions in the WP once the branch is resolved.
Obtaining traces with WP instructions is very challenging as most tracing utilities only annotate committed instructions because they are obtained via binary instrumentation~\cite{luk2005pin}.

We propose using an execution-driven simulator to capture wrong-path instructions
and encode them in the trace. Additionally, we 
suggest several modifications to a trace-driven simulator to support the execution and squashing of wrong-path instructions.

\subsection{WP Trace}
We propose a new trace format to encode a stream of instructions following a mis-speculation, either due to branch misprediction or load-store disambiguation failure.
WP instructions executed before the redirection are discarded or squashed. These instructions are encoded in the trace following the mis-speculated
instruction using additional fields to differentiate them from correct path instructions.
In a modern core with a decoupled front-end, the branch predictor runs ahead of fetch to prefetch instructions in the predicted path. The Branch Predictor Unit (BPU) and Instruction Fetch Unit (IFU) communicate using a Fetch Target Queue (FTQ). When the IFU is redirected on a mis-speculation the FTQ is flushed. The FTQ entries that are flushed are also on the wrong-path. The prefetcher would have issued the FTQ entries so they contribute towards the Instruction Cache pollution in the wrong-path. 
We refer to this as FTQ Prefetch.
Since it is not possible to obtain wrong-path instructions with existing tracing utilities on real CPUs,
we use gem5~\cite{gem5} to generate traces with wrong-path instructions. gem5 is an
execution driven simulator so execution of the WP is modeled in an OOO CPU model.

\subsection{Implementing WP in a Trace Driven Simulator}
In typical trace driven simulators, such as ChampSim, when a branch instruction is encountered, the BPU is queried for the branch direction and target.
The trace, which includes the correct branch target, is then compared with the predicted target to detect mispredictions.
The front-end is stalled till the branch is resolved. Once the branch is resolved
a constant penalty is paid to account for pipeline repair cost and then fetch
is resumed again to process stream of instructions in the correct path. Since wrong-path
instructions are not seen in the pipeline, there is no need to squash or repair any
structures in the pipeline, thus no pollution in the caches.

To model WP instructions in a trace-driven simulator, all stages in the pipeline are modified as follows.
The fetch stage is modified to continue streaming instructions from the trace after
a misprediction.
WP fetch is stopped either when a resteer signal is received, indicating the resolution of the mispeculation.  Resteers may come from the decode stage for unconditional branches or from the execute stage for other types of mispredictions.

In the decode stage unconditional direct branches that are mispredicted are identified.
Once the mispredicted branch is found, all newer (younger) instructions following the
mispredicted branch are squashed. The resteer signal to fetch stage is sent and FTQ,
Decode and Fetch buffers are flushed. Furthermore, the trace until the correct target instruction is skipped.

The execute stage handles all other mispredictions. At execute stage, instructions are executed
out of order from the Reorder Buffer (ROB). The ROB contains instructions in the program
order (in-order). Instructions in the ROB following a mis-speculating instruction are only
in the WP. Once the mis-speculation is resolved all following instructions in
the ROB are flushed. All pipeline structures leading to execute stage are flushed. Rename maps are repaired to remove stale dependencies
introduced by the renamed WP instructions.

\section{Methodology}
\label{sec:methodology}

\subsection{Simulation Infrastructure}
We extended ChampSim~\cite{ChampSim} as described above. 
The CPU core microarchitecture
parameters (Table~\ref{tab:processor_configurations}) are modeled after Intel's Golden Cove core, (aka Alder Lake).
Initially, the CPU is warmed up using 10 million instructions and, the simulation executes 100 million instructions in detailed mode (O3CPU).

We used the gem5~\cite{gem5} simulator to obtain the traces with WP instructions. Traces are obtained
using the detailed O3CPU with same core parameters in Table~\ref{tab:processor_configurations}. The gem5
simulator also models the FDIP pipeline.

\subsection{Benchmarks}
\label{sec:benchmarks}
We examined 14 widely-used server applications from various benchmark suites ~\cite{spec2017, java_renaissance, oltp_bench, kasture2016tailbench, cloudsuite, chipyard, browser_bench} as depicted on Table \ref{tab:benchmarks}. 
These selected workloads contains traditional SPEC'17 CPU~\cite{spec2017} and datacenter workloads with large code footprints. 

\begin{table}[ht]
\begin{tabular}{|c|l|}
\hline
Front-End & \begin{tabular}[c]{@{}l@{}}24 FTQ entries, 16K entries, 8-way BTB\\ 64 KB TAGE~\cite{tage}/ITTAGE~\cite{ittage} branch predictors,\\ 32KB, 8-way L1I with latency of 2 cycles\end{tabular} \\ \hline
Execution & \begin{tabular}[c]{@{}l@{}}512/194/144/112 ROB/Issue/Load/Store entries,\\ 12 wide Decode and Retire, 448/400 Int/Vec. Registes\end{tabular} \\ \hline
Caches & \begin{tabular}[c]{@{}l@{}}64KB, 16-way L1D with latency of 1 cycle,\\ 1 MB, 16-way Private L2C with latency of 10 cycles,\\ 2 MB, 16-way Shared LLC with latency of 20 cycles\end{tabular} \\ \hline
\end{tabular}
\vspace{1pt}
\caption{Processor configurations for ARM ISA}
\vspace{-10pt}
\label{tab:processor_configurations}
\end{table}

\vspace{-15pt} 
\begin{table}[ht]
\centering
\begin{tabular}{|ll}
\multicolumn{1}{c}{Benchmark Suite}   & \multicolumn{1}{c}{Benchmarks}     \\ \hline
\multicolumn{1}{|l|}{SPEC~\cite{spec2017}}            & \multicolumn{1}{l|}{401.bzip2, 429.mcf, 500.perlbench, 523.xalancbmk} \\
\multicolumn{1}{|l|}{} & \multicolumn{1}{l|}{514.leela, 520.omentpp, 531.deepsjeng, 557.xz} \\ \hline
\multicolumn{1}{|l|}{DaCapo~\cite{dacapo_bench}}          & \multicolumn{1}{l|}{cassandra, kafka, tomcat} \\ \hline
\multicolumn{1}{|l|}{Renaissance~\cite{java_renaissance}}     & \multicolumn{1}{l|}{finagle-chirper, finagle-http} \\ \hline
\multicolumn{1}{|l|}{OLTB Bench~\cite{oltp_bench}}      & \multicolumn{1}{l|}{tpcc, wikipedia} \\ \hline
\multicolumn{1}{|l|}{Tailbench~\cite{kasture2016tailbench}}       & \multicolumn{1}{l|}{specjbb, web-search, xapian} \\ \hline
\multicolumn{1}{|l|}{Cloudsuite V4~\cite{cloudsuite}}   & \multicolumn{1}{l|}{data-serving, media-streaming} \\ \hline
\multicolumn{1}{|l|}{Chipyard~\cite{chipyard}}        & \multicolumn{1}{l|}{verilator} \\ \hline
\multicolumn{1}{|l|}{Browser Bench~\cite{browser_bench}}   & \multicolumn{1}{l|}{speedometer2.0} \\ \hline
\end{tabular}
\vspace{1pt}
\caption{Benchmarks used to evaluate}
\vspace{-25pt}
\label{tab:benchmarks}
\end{table}

\todo{How close are they with gem5? }

\section{Results}
\label{sec:results}
In this section, we present the performance analysis of simulating WP execution known as WP mode and compare it with CP mode that runs only CP instructions ignoring  all WP effects on all benchmarks detailed in Section IV-B. Furthermore, we examine the impact of WP mode on the cache hierarchy, providing measurements for the L1 instruction cache (L1I), L1 data cache (L1D), L2 cache, and L3 cache. These insights offer a comprehensive understanding of the implications of WP execution on cache performance across different levels of the cache subsystem.

\subsection{WP Impact on Instructions processed}
\begin{figure}[ht]
    \centering
    \vspace{-10pt}
    \includegraphics[width=\columnwidth]{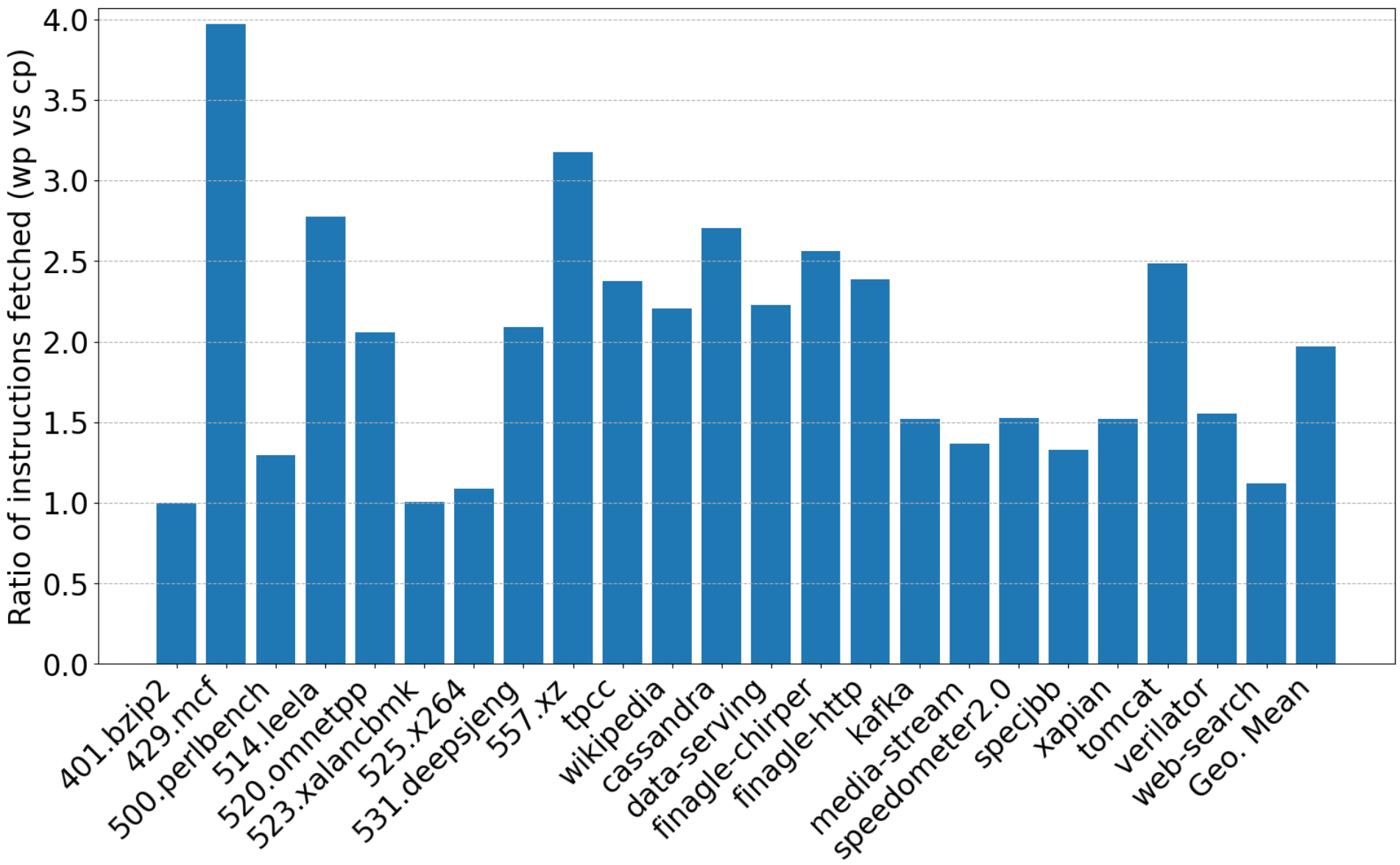}
    \vspace{-15pt}
    \caption{Relative Increase in Instructions in WP vs CP}
    \vspace{-15pt}
    \label{fig:wp_insts}
    
\end{figure}

Modern processors have aggressive decoupled front ends, large structures, and deep ROBs, allowing a substantial number of WP instructions to be fetched and executed before a branch misprediction is identified and corrected. Fig \ref{fig:wp_insts} shows that on average, 83\% more instructions are fetched in WP mode, with many benchmarks such as \texttt{429.mcf}, \texttt{514.leela}, and \texttt{finagle-chirper} showing greater than a 2.5-fold increase. This increase strongly correlates with branch misprediction rates, which control how often we proceed down the wrong path, and structural factors like FTQ size and ROB size, determining how deep we go down the wrong path. ROB occupancy at the time of misprediction in WP is, on average, close to 2 times that seen on the CP. This also leads to considerably more ROB Full events in WP mode. Benchmarks with higher misprediction rates and lower ROB occupancy show the greatest increases in WP instructions executed. 

\subsection{Cache State}
\begin{figure}[ht]
  \centering
  \vspace{-10pt}
  \includegraphics[width=\columnwidth]{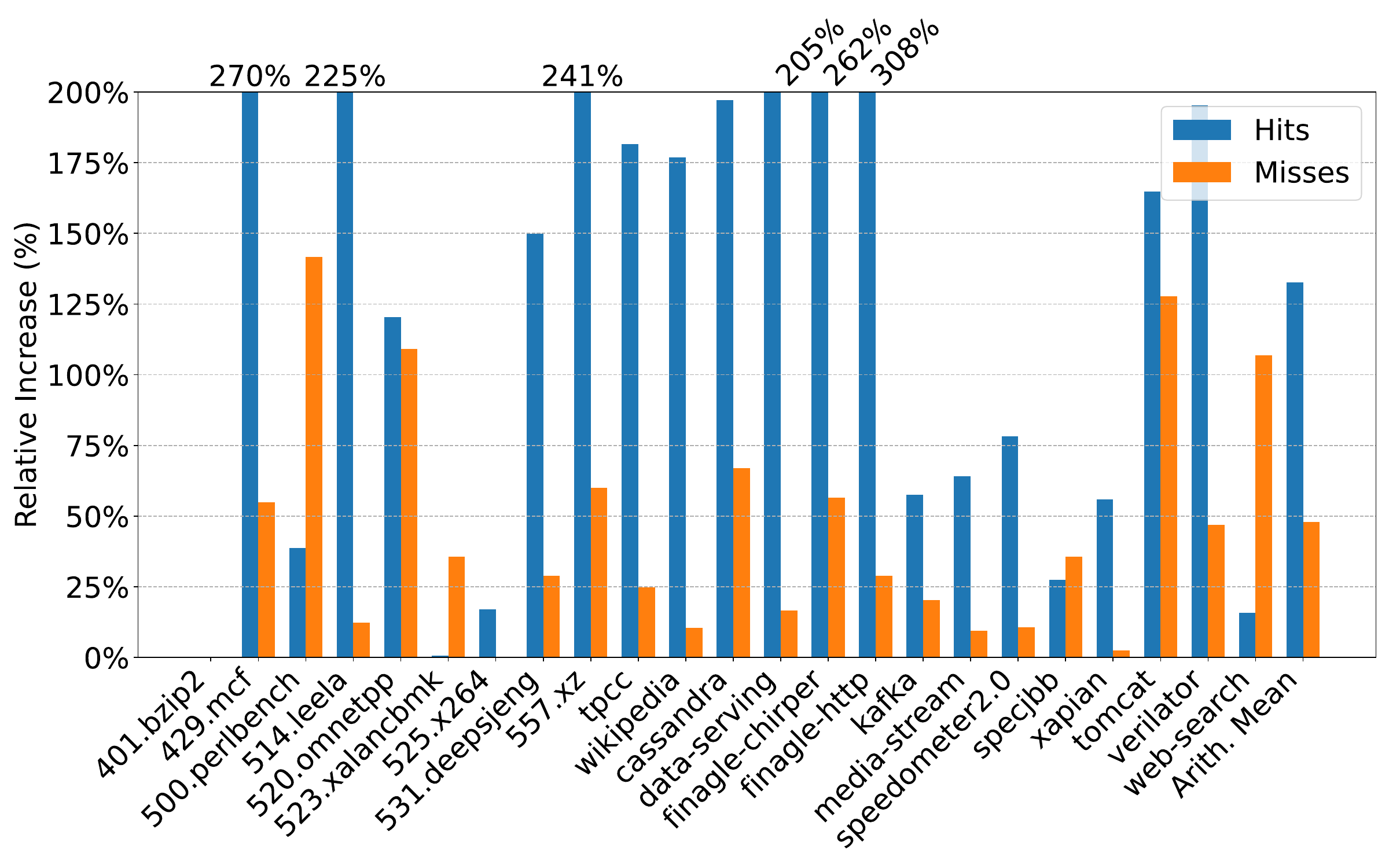}
  \vspace{-25pt}
  \caption{Cache stats for WP and CP modes in L1I}
  \vspace{-4pt}
  \label{fig:l1I_stats}
\end{figure}

\begin{figure}[ht]
  \centering
  \vspace{-10pt}
  \includegraphics[width=\columnwidth]{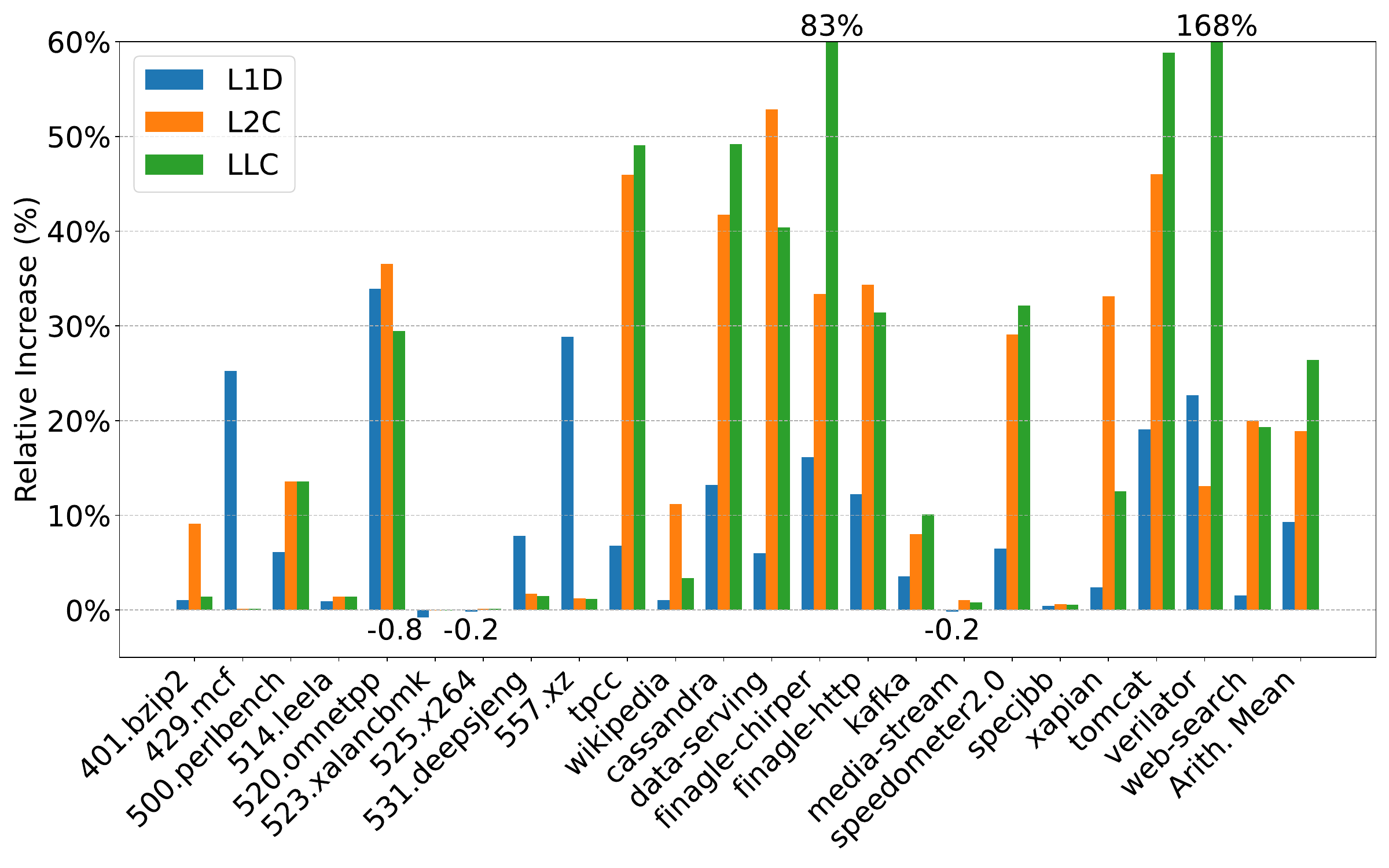}
  \vspace{-15pt}
  \caption{Cache miss stats for WP and CP modes in for all caches}
  \vspace{-10pt}

  \label{fig:miss_rest}
\end{figure}

\begin{figure}[ht]
  \centering
  \vspace{-10pt}
  \includegraphics[width=\columnwidth]{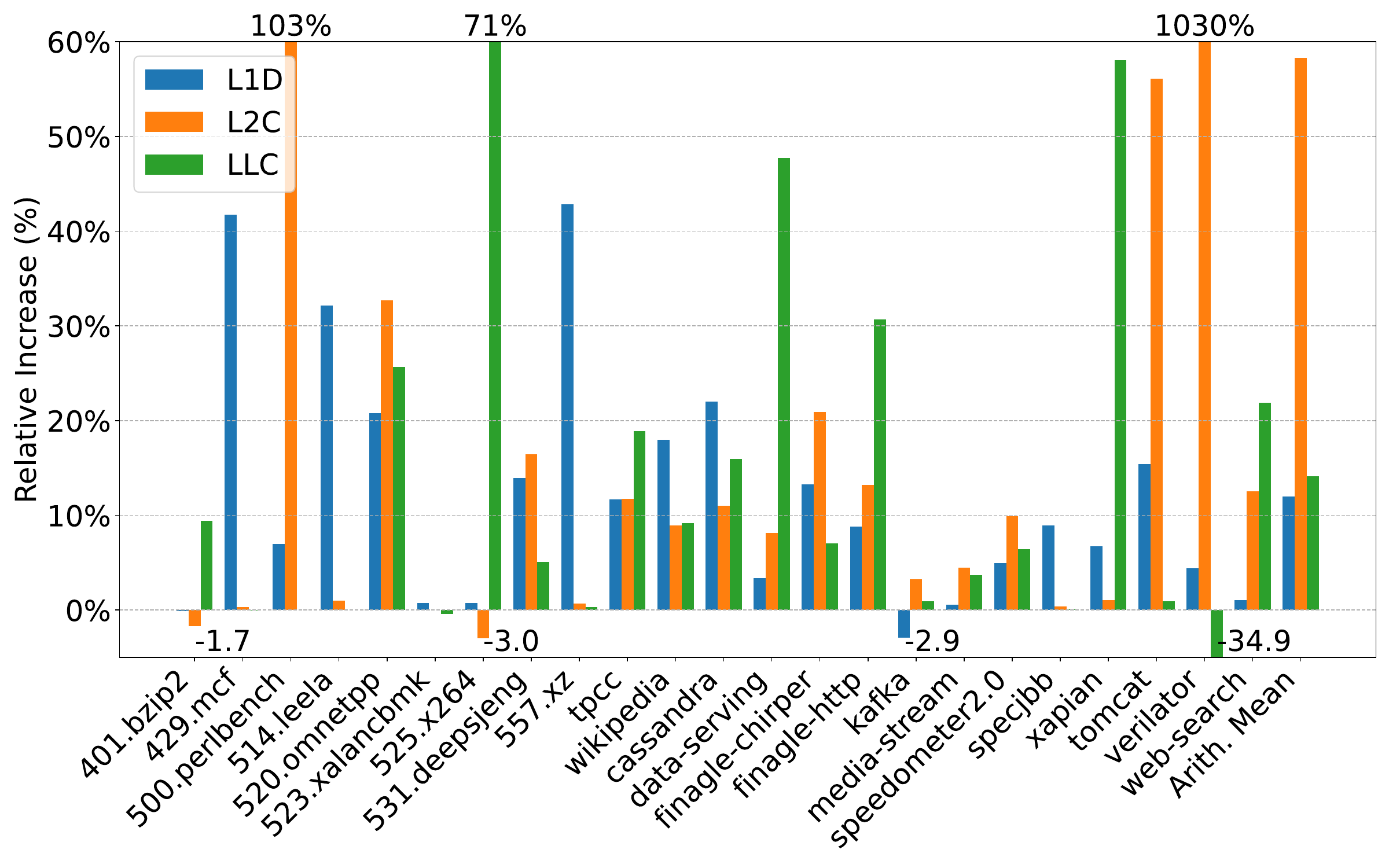}
  \vspace{-15pt}
  \caption{Relative increase in cache hits of WP mode w.r.t CP mode}
  \vspace{-10pt}
  \label{fig:hits_rest}
\end{figure}

Modeling the execution of WP instructions has a direct impact on cache state. L1I cache is the most transformed by the excess WP instructions. Fig \ref{fig:l1I_stats} shows that, on average, there is a 32\% increase in misses in the L1I cache in WP mode compared to CP mode. Notably, four benchmarks more than double the number of misses, and the \texttt{500.perlbench} benchmark exhibits a 147\% increase. An even more pronounced effect is observed in the hit patterns of the L1I cache. In WP mode, there is an average increase of 66\% in hits, with over half the benchmarks showing increases greater than 150\%. The cache states of the other caches are also affected by WP. Fig \ref{fig:miss_rest} shows, on average, L1D misses increase by 9.3\%, L2C misses increase by 18.5\% and LLC misses increase by 26\%. Fig \ref{fig:hits_rest}  shows, on average, L1D hits increases by 12\%, L2C hits increase by 58.3\% and LLC hits increase by 26.3\%. Thus WP also affects the data caches and must not be ignored. The pressure on the L2C cache correlates more closely with performance than the L1D cache. 
Fundamentally, the access patterns of caches differ significantly when WP is enabled. Furthermore metadata used by cache policies like reuse percentage, stride and other temporal and spatial patterns are also different. Thus any work impacted by cache behaviour is incomplete without modelling WP accurately. The higher hit rates may ensure that high-reuse lines are preserved longer in the design, but they may also lead to the pollution of metadata updates. Depending on the nature of the benchmark, this effect could positively or negatively affect performance at different phases of the program.

\begin{figure}[ht]
  \centering
  \includegraphics[width=\columnwidth]{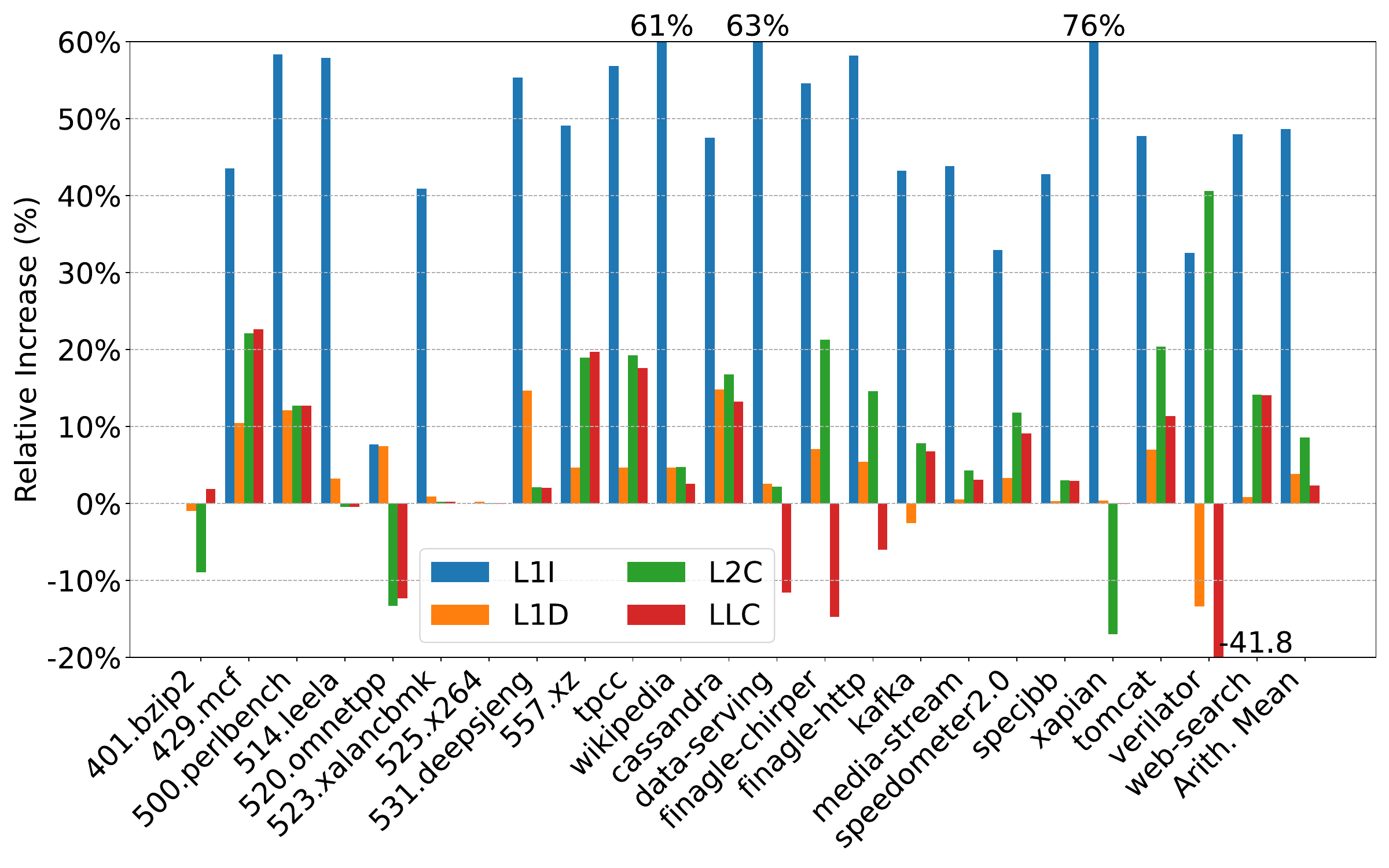}
  \caption{Reduction of cache misses in the CP in WP mode}
  \vspace{-10pt}
  \label{fig:reduction_in_cp_misses}
\end{figure}

\subsection{WP Impact on Cache State}
The cache lines that were brought in due to WP execution can be useful or useless. Wrong-path lines that are eventually used on the correct path can enhance performance and is considered WP a useful fill, leading to a positive impact like a prefetch. Conversely, lines that are evicted without being used can degrade performance and is considered WP a useless fill, resulting in a negative impact. 
WP-induced useless fills reduce effective occupancy of the caches and may also evict other critical useful lines, thus leading to lost performance.  However, WP-induced useful fills help in prefetching lines accessed in later by the CP, resulting in lower cache misses.  On average 67\% of WP fills in L1I cache are useful. Thus we see a strong prefetching effect. Further Fig \ref{fig:reduction_in_cp_misses} shows the reduction in misses along the CP when running in WP mode. L1I sees a reduction of 44\% in the misses along the correct path, with L1D showing 3.85\% and L2C showing 8.58\% reduction. Although the overall misses increase, the prefetching effect of WP reduces the number of misses along CP across instruction and data caches. 

\subsection{Performance Impact}
\begin{figure}[ht]
    \centering
    \vspace{-10pt}
  \includegraphics[width=\columnwidth]{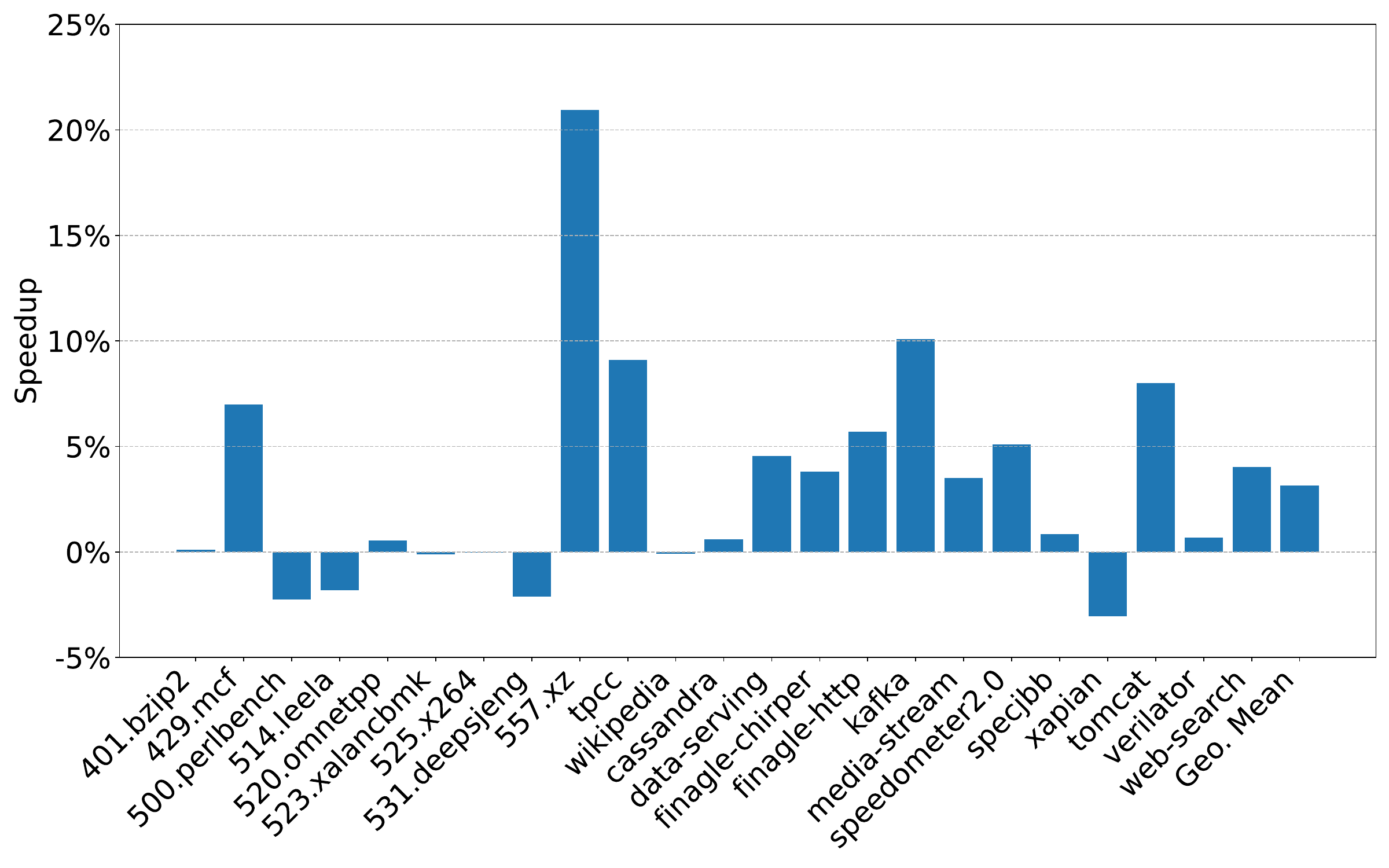}
    \vspace{-15pt}
    \caption{IPC improvement with of WP mode w.r.t CP mode}
    \vspace{-15pt}    
    \label{fig:ipc}
\end{figure}

As discussed in the previous sections, the performance impact of WP cannot be approximated by considering cache behavior, branch mispredictions, or back-end impacts in isolation. 
WP influences these factors differently across various parts of the benchmarks, and only a comprehensive analysis of all these effects together can adequately capture the true impact of WP. Figure~\ref{fig:ipc} shows the IPC speedup of WP mode over CP mode and the branch MPKI of all benchmarks. Benchmarks like \texttt{tomcat} and \texttt{cassandra} have high branch MPKI. In contrast, \texttt{523.xalancbmk,} and \texttt{525.x264} have low branch MPKI so it has negligible variance with WP. The expected trend is that the higher the branch misprediction rate is, the higher the number of WP instructions. However, in \texttt{verilator} the branch MPKI is 9.14 the WP instructions are only 55\%(lower than other benchmarks with similar MPKI). This is because \texttt{verilator} has high number of unconditional branches which are resolved in the decode stage. Thus instructions fetched in the WP are squashed even before they reach decode stage. 

The prefetching effect of WP dominates in majority of benchmarks thus WP showing positive impact in majority of benchmarks with a speedup of upto 20.9\% and a mean gain of 3.26\%.  WP can also have negative impact which resulted in performance loss in seven benchmarks with \texttt{xapian} showing the minimum of -3.09\% speeedup.

\todo{WP instructions need not always result in cache fills this
is one major difference with respect to \cite{wrong_path_intel}.} 

\section{Conclusion}
Complex modern OOO CPUs have speculative pipeline which have significant
impact on the state of cache. The effect of WP execution is dependent
on various characteristics of benchmarks. The impact of WP is often
ignored by fast less accurate trace-driven simulators. We provide a model
to simulate WP in trace-drive simulator with a new trace format
to encode WP instructions. As the traces are generated from a execution-driven simulator, the advantage of correct WP is achieved still at the quick speed of a trace-driven simulator. The WP enabled traces are robust
enough to capture various characteristics of benchmarks and CPU parameters.
\vspace{-5pt}

